# Accurate Effective Medium Theory for the Analysis of Spoof Localized Surface Plasmons in Textured Metallic Cylinders

Babak Rahmani, Amirmasood Bagheri, and Amin Khavasi

The authors are with the Electrical Engineering Department, Sharif University of Technology, Tehran 11155-4363, Iran

Email: khavasi@sharif.edu

*Abstract*

It has been recently demonstrated that textured closed surfaces which are made out of perfect electric conductors (PECs) can mimic highly localized surface plasmons (LSPs). Here, we propose an effective medium which can accurately model LSP resonances in a two-dimensional periodically decorated PEC cylinder. The accuracy of previous models is limited to structures with deep-subwavelength and high number of grooves. However, we show that our model can successfully predict the ultra-sharp LSP resonances which exist in structures with relatively lower number of grooves. Such resonances are not correctly predictable with previous models that give some spurious resonances. The success of the proposed model is indebted to the incorporation of an effective surface conductivity which is created at the interface of the cylinder and the homogeneous medium surrounding the structure. This surface conductivity models the effect of higher diffracted orders which are excited in the periodic structure. The validity of the proposed model is verified by full-wave simulations.

## I. Introduction

High confinement and enhanced electromagnetic fields are among significant characteristics of surface plasmons (SPs) at optical frequencies which have stimulated great research interest over the last decades [1], [2]. It is a well-known fact that SPs can be categorized into two main types. The first type is the propagating modes confined to the dielectric-metal interfaces and the second type is localized surface plasmons (LSPs) which exist in subwavelength particles. The exquisite characteristics of LSPs can be tailored to bring forth a plethora of applications including plasmonic antennas [3], surface-enhanced Raman scattering [4], photovoltaics [5], hyperlenses [6] and bio-sensors [7].

The promising applications of plasmonics in optical frequencies have spurred scientists to find a way to mimic SPs at lower frequencies. In the pioneering work of Pendry et. al. in 2004, the authors showed that textured metallic structures can support subwavelength resonances resembling SPs at optical frequencies [8]. These collective oscillations, the so-called spoof surface plasmons (SSPs), can be emulated in microwave (MW) and terahertz (THz) frequency regimes by means of periodic metallic surfaces [8]. SSPs have opened up new horizons which enable designing metamaterials with properties otherwise impossible to achieve. Geometrically manipulated media for achieving high positive refractive indices [9] and miniaturization of THz devices [10], [11] are among the vast possible applications of SSP-assisted resonances.

It has been recently demonstrated that closed-surface metallic structures can support electromagnetic resonances in MW and THz frequencies which are very similar to LSPs at optical regime [12]. Pors et. al. named these resonances as spoof-LSPs. Following this work, a manifold of structures supporting spoof-LSPs have been presented. In [13], a two-dimensional (2D) periodically decorated cylinder with multiple groove depths has been suggested which can excite multi-band LSP resonances. Likewise, stimulation of higher order LSPs in textured metallic disks [14], [15] and subwavelength high contrast 2D cylindrical gratings [16] have been reported. Other structures supporting spoof LSP are non-concentric closed surfaces [17], corrugated ring resonators [18] and textured open metal surfaces [19]. In another work, the elongated periodically textured cylinders have been proposed which support magnetic LSPs [20]. Yet, it should be emphasized that deep-subwavelength groove widths are the common characteristic of all textured closed-surfaces introduced so far.

In this article, we aim to go beyond this limitation by introducing an effective medium theory which is able to accurately predict highly confined spoof-LSP resonances in the far-infrared regime for a 2D periodically decorated cylinder with relatively lower number of grooves as compared with previous structures [12]-[20]. For the analysis of the structure, we derive an analytical effective medium model which can successfully predict the ultra-sharp resonances. It is of the utmost importance to note that the structure, due to the presence of higher diffracted orders associated with lower number of grooves, cannot be accurately modeled by the conventional effective medium which has been widely used in the previous works [12]. Instead, we demonstrate that by incorporating an effective surface conductivity into our effective medium model, the features of spoof-LSP resonances in the structure can be captured much more accurately even if we are not in deep-subwavelength regime.

The paper is organized as follows: In section II, we propose an effective medium model for a 2D periodically textured cylinder for TM polarization. Numerical examples are given in section III to show the validity of the proposed model. Finally, conclusions are drawn in section IV. A time dependence of the form $e^{j\omega t}$ is assumed throughout this paper.

## II. DERIVATION OF THE EFFECTIVE MEDIUM

In this section, we use the effective medium theory to propose an equivalent model for the analysis of a 2D textured perfect electric conductor (PEC) cylinder. The schematic of the structure is depicted in Fig. 1.

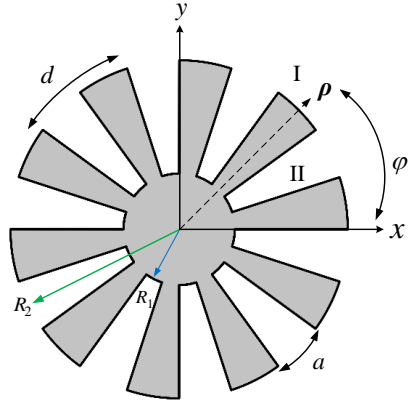

Fig. 1 A 2D textured PEC cylinder with period $d$ and groove height and width $h=R_2-R_1$ and $a$, respectively. Region I ($r > R_2$) is the homogeneous medium, and region II ($R_1 < r < R_2$) is the grooves region.

The structure is periodic in the $\varphi$-direction with the periodicity of $d = 2\pi R_2 / N$ at the outer interface, where $N$ is the total number of grooves and $R_2$ is the outer radius of the cylinder. The width and height of the grooves are $a$ (at the outer interface) and $h = R_2 - R_1$, respectively. For the sake of simplicity, the homogeneous media around the cylinder (region I) and inside the grooves (region II) is considered to be free space.

We assume that the structure depicted in Fig. 1 is illuminated by a TM polarized plane wave, i.e. $H_z^{inc} = H_0 e^{-j\mathbf{k}_i \cdot \mathbf{r}}$, where $\mathbf{k}_i = k_0 \cos(\varphi_i)\hat{x} + k_0 \sin(\varphi_i)\hat{y}$ is the incident wavevector, $\boldsymbol{\rho} = \rho\cos(\varphi)\hat{x} + \rho\sin(\varphi)\hat{y}$ and $H_0$ is the amplitude of the incident wave. Also, $k_0$ represents the free space wavenumber. For the simplicity's sake, we choose $\varphi_i = 0$ in

$k_i$ expression. The incident wave can be expanded in a series of Bessel functions of the first kind and order $m$:

$$H_z^{inc}(\rho,\varphi) = \sum_{m=-\infty}^{\infty} A_m J_m(k_0\rho) e^{-jm\varphi} \tag{1}$$

where $A_m$ is the expansion coefficient which is equal to $H_0(-j)^m$. Thus, the tangential electric and magnetic fields in region I can be written as [21]

$$H_z^{(I)}(\rho,\varphi) = \sum_{m=-\infty}^{\infty} A_m J_m(k_0\rho) e^{-jm\varphi} + \sum_{m=-\infty}^{\infty}\sum_{n=-\infty}^{\infty} B_{mn} H_{v_{mn}}^{(2)}(k_0\rho) e^{-jv_{mn}\varphi} \tag{2}$$

and

$$E_\varphi^{(I)}(\rho,\varphi) = \sum_{m=-\infty}^{\infty} A_m \frac{-k_0}{j\omega\varepsilon_0} J_m'(k_0\rho) e^{-jm\varphi} + \sum_{m=-\infty}^{\infty}\sum_{n=-\infty}^{\infty} B_{mn} \frac{-k_0}{j\omega\varepsilon_0} H_{v_{mn}}^{(2)'}(k_0\rho) e^{-jv_{mn}\varphi} \tag{3}$$

where $v_{mn}$ is the wavevector component along the $\varphi$-direction defined as

$$v_{mn} = m + nN, \quad (m,n = 0, \pm 1, \pm 2,...) \tag{4}$$

One the other hand, the electromagnetic fields in region II can be written as follows [12]

$$H_z^{(II)}(\rho,\varphi) = \sum_{l=0}^{\infty} \cos(l\alpha\varphi)[C_l H_{l\alpha}^{(2)}(k_0\rho) + D_l H_{l\alpha}^{(1)}(k_0\rho)] \tag{5}$$

and

$$E_\varphi^{(II)}(\rho,\varphi) = \sum_{l=0}^{\infty} \frac{-k_0}{j\omega\varepsilon_0} \cos(l\alpha\varphi)[C_l H_{l\alpha}^{(2)'}(k_0\rho) + D_l H_{l\alpha}^{(1)'}(k_0\rho)] \tag{6}$$

where $\alpha = R_2\pi/a$. In the aforementioned expressions, $H_{v_{mn}}^{(1)}(.)$ and $H_{v_{mn}}^{(2)}(.)$ are the Hankel functions of the first and second type and order $v_{mn}$.

By applying the continuity conditions of tangential electromagnetic fields for a given $m$ at the interface between regions I and II, the electromagnetic field distributions in both regions can be easily obtained. First, multiplying the electric fields by $e^{jv_{mp}\varphi}$ and integrating both sides over one period at $r = R_2$, we find:

$$B_{mp} H_{v_{mp}}^{(2)'}(k_0 R_2) + A_m J_m'(k_0 R_2)\delta_{p0} = C_l \psi P_{mpl}^+ \frac{R_2}{d} \tag{7}$$

where $\delta_{p0}$ is the Kronecker delta function. Moreover, $\psi$ and $P_{mpl}^+$ are defined as follows

$$\psi = [H_{l\alpha}^{(2)'}(k_0 R_2) - \frac{H_{l\alpha}^{(2)'}(k_0 R_1)}{H_{l\alpha}^{(1)'}(k_0 R_1)} H_{l\alpha}^{(1)'}(k_0 R_2)] \qquad (8)$$

$$P_{mpl}^{\pm} = \int_0^{a/R_2} \cos(l\alpha\varphi) e^{\pm j v_{mp} \varphi} d\varphi \qquad (9)$$

It should be noted that the condition which requires the tangential electric field to vanish at the internal PEC interface $r = R_1$, i.e. $D_l = -C_l H_{l\alpha}^{(2)'}(k_0 R_1) / H_{l\alpha}^{(1)'}(k_0 R_1)$, has been applied in (7).

Second, multiplying the magnetic fields by $\cos(l\alpha\varphi)$ and integrating both sides over one groove width at $r = R_2$, we find:

$$\sum_{n=-\infty}^{\infty} B_{mn} H_{v_{mn}}^{(2)}(k_0 R_2) P_{mnl}^{-} + A_m J_m(k_0 R_2) P_{m0l}^{-}$$

$$= C_l \chi \frac{a}{R_2} U_l \qquad (10)$$

where $U_0 = 1$ for $l = 0$, $U_l = 1/2$ for $l > 0$ and

$$\chi = [H_{l\alpha}^{(2)}(k_0 R_2) - \frac{H_{l\alpha}^{(2)'}(k_0 R_1)}{H_{l\alpha}^{(1)'}(k_0 R_1)} H_{l\alpha}^{(1)}(k_0 R_2)] \qquad (11)$$

By rearranging (7)-(11), one can solve for the reflection coefficients $B_{mp}$. These coefficients have to be obtained by matrix calculations. However, by considering only the principal mode inside the grooves (*l*=0), the reflection coefficients can be written in a more simplified form

$$B_{mp} = \frac{\psi P_{mp0}^{+} P_{m00}^{-} \phi}{H_{v_{mp}}^{(2)'}(k_0 R_2)[\chi \frac{ad}{R_2^2} - \psi \sum_n |P_{mn0}^{+}|^2 \frac{H_{v_{mn}}^{(2)}(k_0 R_2)}{H_{v_{mn}}^{(2)'}(k_0 R_2)}]} +$$

$$- \delta_{p0} \frac{J_m'(k_0 R_2)}{H_{v_{m0}}^{(2)'}(k_0 R_2)} \qquad (12)$$

where $\phi$ is defined as follows

$$\phi = [J_m(k_0 R_2) - \frac{H_{v_{m0}}^{(2)}(k_0 R_2)}{H_{v_{m0}}^{(2)'}(k_0 R_2)} J_m'(k_0 R_2)] \qquad (13)$$

It is of particular interest to note that (12) resembles the reflection coefficient of the homogeneous effective medium model depicted in Fig. 2.

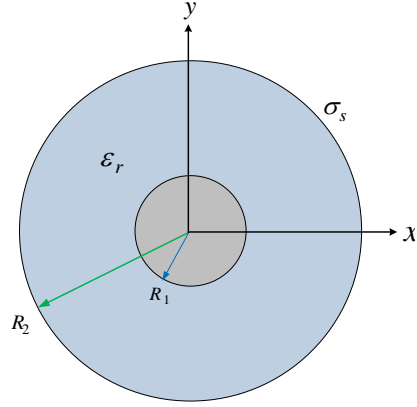

Fig. 2 The proposed effective medium model for the structure shown in Fig 1.

The *m*th reflection coefficient of the effective medium model is defined as follows:

$$\rho_m = \frac{\phi \frac{\overline{\psi}}{\sqrt{\varepsilon_r}}}{H_m^{(2)'}(k_0 R_2)[\overline{\chi} - \frac{\overline{\psi}}{\sqrt{\varepsilon_r}}(j\sigma_s \eta_0 + \frac{H_m^{(2)}(k_0 R_2)}{H_m^{(2)'}(k_0 R_2)})]} - \frac{J_m'(k_0 R_2)}{H_m^{(2)'}(k_0 R_2)} \quad (14)$$

where

$$\overline{\psi} = [H_m^{(2)'}(\kappa R_2) - \frac{H_m^{(2)'}(\kappa R_1)}{H_m^{(1)'}(\kappa R_1)} H_m^{(1)'}(\kappa R_2)] \quad (15)$$

and

$$\overline{\chi} = [H_m^{(2)}(\kappa R_2) - \frac{H_m^{(2)'}(\kappa R_1)}{H_m^{(1)'}(\kappa R_1)} H_m^{(1)}(\kappa R_2)] \quad (16)$$

in which, $\eta_0$ is the free space impedance and $\kappa = k_0 \sqrt{\varepsilon_r}$. Comparing $B_{m0}$ from (12) and $\rho_m$ from (14), we can easily determine the sought-after effective permittivity and surface conductivity of the proposed equivalent model for a given *m* as follows

$$\varepsilon_r = (\frac{ad / R_2^2}{|P_{m00}^+|^2})^2 \quad (17)$$

$$\sigma_s = -j \frac{\sqrt{\varepsilon_r}}{\eta_0} [\frac{\overline{\chi}}{\overline{\psi}} - \frac{\chi}{\psi} + \frac{R_2^2}{ad} \sum_{n \neq 0} |P_{mn0}^+|^2 \frac{H_{v_{mn}}^{(2)}(k_0 R_2)}{H_{v_{mn}}^{(2)'}(k_0 R_2)}] \quad (18)$$

It should be noticed that although $\varepsilon_r$ depends on *m*, it can be well approximated by $\varepsilon_r \approx (d/a)^2$ which still provides highly accurate results.

## III. NUMERICAL RESULTS

In this section, we demonstrate the accuracy of our model through several numerical examples. To do so, the proposed effective medium model is compared against the original structure in terms of scattering. The scattering cross section (SCS) of the equivalent model is defined as

$$C_{sca} = \frac{4}{k_0} \sum_{m=-\infty}^{\infty} |\rho_m|^2 \qquad (19)$$

It should be mentioned that full-wave simulations are performed with the commercial simulator COMSOL Multiphysics. As the first example, we assume that the structure is illuminated by a TM polarized plane wave propagating in the $-x$-direction. Figure 3 shows the SCS which has been normalized to the diameter $2R_2$ (normalized SCS) of the structure with geometrical parameters $R_2 = 300\,\mu\text{m}$, $R_1 = 0.33R_2$, $a = 0.4d$, $N = 40$ versus the normalized frequency $R_2/\lambda$ calculated by the conventional effective medium [12] (dashed-dotted line), the here-proposed effective medium (dashed line), and the full-wave simulation (solid line).

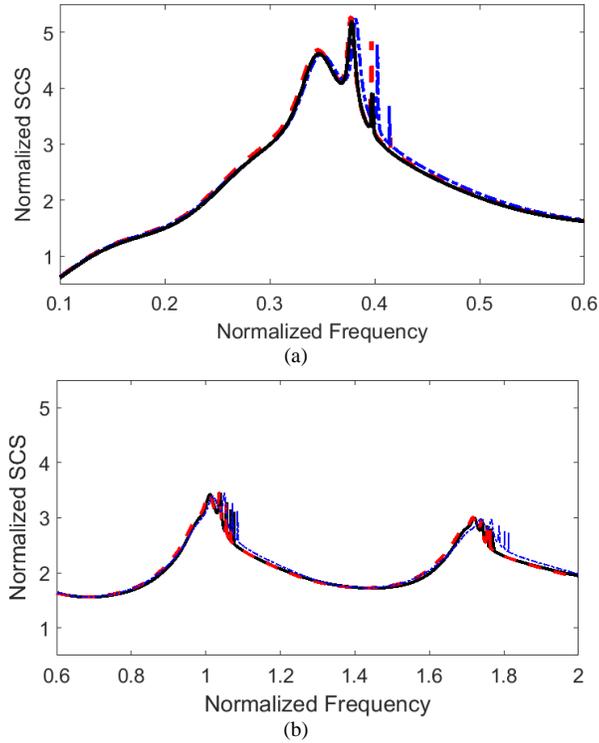

Fig. 3 The normalized SCS versus the normalized frequency for the structure with geometrical parameters $R_2 = 300\,\mu\text{m}$, $R_1 = 0.33R_2$, $a = 0.4d$, $N = 40$ when illuminated by a TM polarized plane wave propagating in the $-x$-direction calculated by the full wave simulations (solid line), the here-proposed effective medium (dashed line) and the conventional effective medium (dashed-dotted line) [12] for intervals (a) $0.1 < R_2/\lambda < 0.6$ and (b) $0.6 < R_2/\lambda < 2$.

Figure 3 clearly demonstrate that the number of resonances in the first and second band is correctly predicted by both the conventional and the here-proposed effective medium. Nevertheless, the results of the conventional effective medium are slightly different from those of the full-wave simulation and the here-proposed effective medium. Remarkably, the latter two methods are in excellent agreement with each other.

By decreasing the total number of grooves in the periodically textured cylinder $(N)$, the discrepancy between the results of the conventional model and those of the full-wave simulations becomes conspicuous. This point is illustrated in the following example.

Consider a textured cylinder with parameters $N = 17$, $R_2 = 300\,\mu\text{m}$, $R_1 = 0.33R_2$, and $a = 0.4d$ which is under illumination of a TM polarized plane wave propagating in $-x$-direction. The normalized SCS versus $R_2/\lambda$ is plotted in Fig. 4.

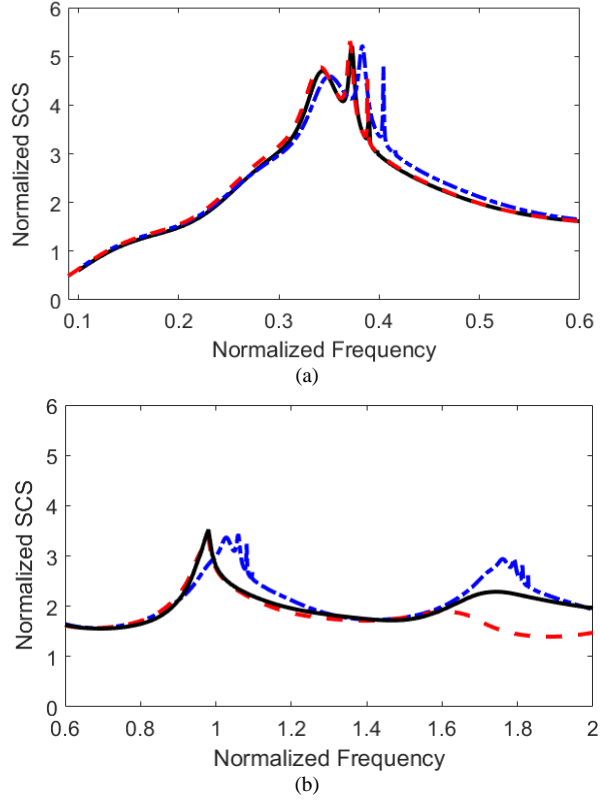

Fig. 4 The normalized SCS versus the normalized frequency for the structure with geometrical parameters $R_2 = 300\,\mu m$, $R_1 = 0.33 R_2$, $a = 0.4 d$, $N = 17$ when illuminated by a TM polarized plane wave propagating in the $-x$-direction calculated by the full wave simulations (solid line), the here-proposed effective medium (dashed line) and the conventional effective medium (dashed-dotted line) [12] for intervals (a) $0.1 < R_2/\lambda < 0.6$ and (b) $0.6 < R_2/\lambda < 2$.

It can be inferred from Fig. 4 that the accuracy of the conventional effective medium is acceptable only in the first band of resonances. By increasing the working frequency, the conventional model becomes erroneous as it fails to correctly predict the number of resonances arising in the second band. In fact, the conventional effective medium suggests existence of spurious resonances. In sheer contrast, the here-proposed effective medium provides very accurate results for both the first and second band of resonances. In other words, it can accurately model the spectral position and the number of resonances emerging in the normalized SCS. The proposed model, however, loses its accuracy in the third band of resonances. In this band, the effect of higher Floquet orders on the scattering must be considered. Therefore, (19) should be modified as follows

$$C_{sca} = \frac{4}{k_0} \sum_{m=-\infty}^{\infty} \sum_{n=-\infty}^{\infty} |B_{mn}|^2 \qquad (20)$$

Due to the homogeneity of the proposed model, higher Floquet orders ($B_{mn}$, $n \neq 0$) cannot be mimicked and thus, their effects on the SCS are not accounted for in (19). These points are demonstrated in Fig. 5.

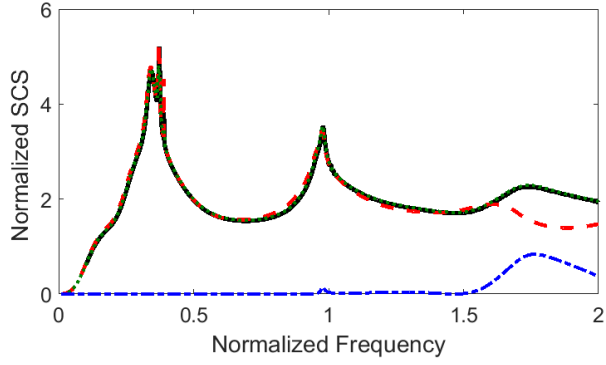

Fig. 5 The normalized SCS versus the normalized frequency for the structure with geometrical parameters $R_2 = 300\,\mu m$, $R_1 = 0.33 R_2$, $a = 0.4d$, $N = 17$ when illuminated by a TM polarized plane wave propagating in the –x-direction calculated by the full wave simulations (solid line), the modal method (dotted line) and the here-proposed effective medium (dashed line). The portion of normalized SCS attributed to the higher Flouqet orders is separately calculated (dashed-dotted line).

Figure 5 depicts the normalized SCS for the previous example using the full-wave simulations, the modal method presented in (7)-(11) and the here-proposed effective medium. Furthermore, the portion of normalized SCS attributed to the higher Floquet orders is separately plotted. Evidently, the proposed model is accurate as long as the power carried by higher Floquet orders is negligible.

The accuracy of the proposed model is further demonstrated in Fig. 6 by plotting the normalized SCS versus the normalized frequency and the number of grooves (*N*) in a structure with parameters $R_2 = 300\,\mu m$, $R_1 = 0.33 R_2$, and $a = 0.4d$.

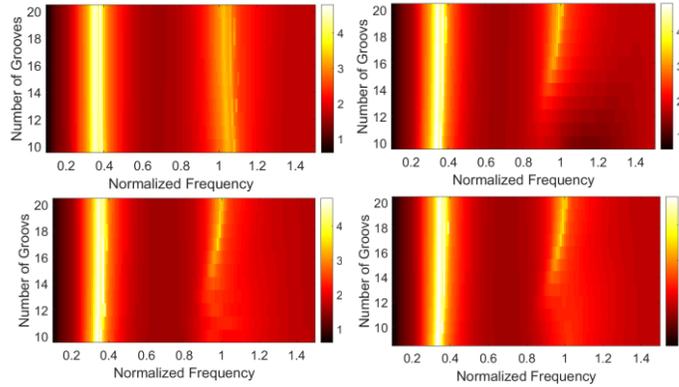

Fig. 6 The normalized SCS versus the normalized frequency and the number of grooves (*N*) for the structure with geometrical parameters $R_2 = 300\,\mu m$, $R_1 = 0.33 R_2$, $a = 0.4d$ when illuminated by a TM polarized plane wave propagating in the –x-direction calculated by the conventional (top left) and the here-proposed effective medium (top right), the full-wave simulations (bottom left) and the modal approach (bottom right).

Finally, we plot the normalized SCS versus the normalized width ($a/d$) in the previous example at $R_2/\lambda = 1$ using the conventional effective medium, the here-proposed model and the modal method. It is obvious from Fig. 7 that the proposed effective medium model still provides fairly accurate results for different groove widths.

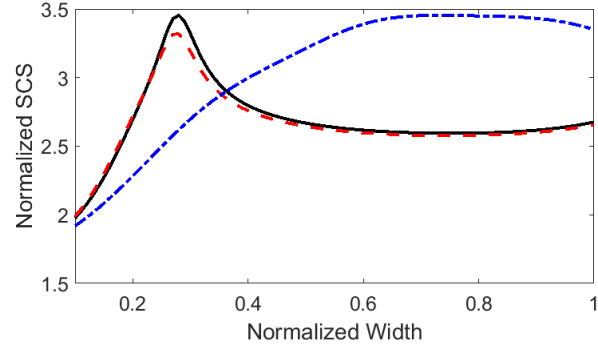

Fig. 7 The normalized SCS versus the normalized width at $R_2/\lambda = 1$ for the structure with geometrical parameters $R_2 = 300\,\mu\text{m}$, $R_1 = 0.33 R_2$, $N = 17$ when illuminated by a TM polarized plane wave propagating in the –x-direction calculated by the modal method (solid line), the here-proposed effective medium (dashed line) and the conventional effective medium (dashed-dotted).

## IV. Conclusion

In summary, we have proposed an effective medium model for a 2D periodically textured PEC cylinder for the TM polarization. The importance of our model stems from the fact that all the previous models fail to mimic the original structure when the total number of grooves in the periodic cylinder substantially decreases. In this case, the conventional effective medium model incorrectly predicts spurious modes. On the contrary, we have shown that by introducing a surface conductivity at the interface of the cylinder and the homogeneous medium surrounding the structure, the number of LSP resonances as well as their spectral position can be successfully predicted. Thus, it can be inferred that ultra-sharp LSPs are not limited to particles with extremely subwavelength grooves.

The limitation of the proposed effective medium model has been also discussed. We showed that the accuracy of the presented model is preserved as long as the power carried by higher Floquet orders is negligible.

The findings of this work may pave the path toward analysis and design of miscellaneous LSP-assisted devices.